\documentclass[twocolumn,preprintnumbers,amsmath,amssymbn,reprint,superscriptaddress]{revtex4}

\usepackage{graphicx}
\usepackage{dcolumn}
\usepackage{bm}
\usepackage{epstopdf}

\usepackage[USenglish]{babel}
\addto{\captionsUSenglish}{}
\usepackage[utf8]{inputenc} 
\usepackage[hidelinks]{hyperref}
\usepackage{amssymb,amsmath}

\newcommand{\ket}[1]{| #1\rangle}

\newcommand*\colvec[3][]{\begin{pmatrix}\ifx\relax#1\relax\else#1\\\fi#2\\#3\end{pmatrix}}
\newcommand{\sinc}[1]{\text{sinc}\left( #1 \right)}

\newcommand{\qq}{\mathbf{q}}

\begin{document}

\title{Quantifying the Momentum Correlation between Two Light Beams by Detecting One}

\author{Armin Hochrainer}
\email{armin.hochrainer@univie.ac.at}
\affiliation{Institute for Quantum Optics and Quantum Information, Austrian Academy of Sciences, Boltzmanngasse 3, Vienna A-1090, Austria.}
\affiliation{Vienna Center for Quantum Science and Technology (VCQ), Faculty of Physics, Boltzmanngasse 5, University of Vienna, Vienna A-1090, Austria.}

\author{Mayukh Lahiri}%
\affiliation{Institute for Quantum Optics and Quantum Information, Austrian Academy of Sciences, Boltzmanngasse 3, Vienna A-1090, Austria.}
\affiliation{Vienna Center for Quantum Science and Technology (VCQ), Faculty of Physics, Boltzmanngasse 5, University of Vienna, Vienna A-1090, Austria.}

\author{Radek Lapkiewicz}
\affiliation{Institute for Quantum Optics and Quantum Information, Austrian Academy of Sciences, Boltzmanngasse 3, Vienna A-1090, Austria.}
\affiliation{Vienna Center for Quantum Science and Technology (VCQ), Faculty of Physics, Boltzmanngasse 5, University of Vienna, Vienna A-1090, Austria.}
\author{Gabriela B. Lemos}
\affiliation{Institute for Quantum Optics and Quantum Information, Austrian Academy of Sciences, Boltzmanngasse 3, Vienna A-1090, Austria.}
\affiliation{Vienna Center for Quantum Science and Technology (VCQ), Faculty of Physics, Boltzmanngasse 5, University of Vienna, Vienna A-1090, Austria.}
\author{Anton Zeilinger}
\email{anton.zeilinger@univie.ac.at}
\affiliation{Institute for Quantum Optics and Quantum Information, Austrian Academy of Sciences, Boltzmanngasse 3, Vienna A-1090, Austria.}
\affiliation{Vienna Center for Quantum Science and Technology (VCQ), Faculty of Physics, Boltzmanngasse 5, University of Vienna, Vienna A-1090, Austria.}

\date{\today}

\begin{abstract}
We report a measurement of the transverse momentum correlation between two photons by detecting only one of them. Our method uses two identical sources in an arrangement, in which the phenomenon of induced coherence without induced emission is observed. In this way, we produce an interference pattern in the superposition of one beam from each source. We quantify the transverse momentum correlation by analyzing the visibility of this pattern. Our approach might be useful for the characterization of correlated photon pair sources and may lead to an experimental measure of continuous variable entanglement, which relies on the detection of only one of two entangled particles.
\end{abstract}

\maketitle

Spatial entanglement \cite{einstein_can_1935} of photon pairs plays an important role in fundamental quantum mechanics \cite{horne1989two,walborn_spatial_2010,schneeloch_introduction_2015}, quantum cryptography \cite{walborn_quantum_2006,walborn_schemes_2008}, quantum teleportation \cite{walborn_quantum_2007} and quantum computation \cite{tasca_continuous-variable_2011}.
A widely used strategy to test spatial entanglement is to directly measure intensity correlations in both near and far fields of the source plane, which are interpreted as correlations in the transverse positions and momenta of the two photons.
Measurements of the transverse momentum correlation between two photons have been performed using a variety of experimental methods \cite{walborn_spatial_2010},
including the scanning of two detectors \cite{pittman_optical_1995,dangelo_identifying_2004}
or two slits \cite{howell_realization_2004}, using detector arrays \cite{osullivan-hale_pixel_2005}, spatial light modulators \cite{hor-meyll_ancilla-assisted_2014},
and cameras capable of resolving individual photon pairs \cite{edgar_imaging_2012}. All of these methods rely on the detection of both of the correlated photons. This fact restricts their applicability to situations where the wavelength of both photons lies in a spectral range, for which sufficiently efficient detectors are available.

If two spatially separated nonlinear crystals emit pairs of photons (signal and idler) by the process of spontaneous parametric down conversion (SPDC) \cite{klyshko_scattering_1969,burnham_observation_1970}, the two resulting signal beams in general do not interfere in lowest order \cite{hong_interference_1988}. This can be understood by the fact that the measurement on an idler photon would provide which-path information about a signal photon. 
However, lowest order interference between the signal beams occurs if the respective idler beams are indistinguishable. 
This phenomenon, known as induced coherence without induced emission, was first observed experimentally in \cite{zou_induced_1991}, following a suggestion by Z. Y. Ou of aligning the two idler beams \footnote{See acknowledgement sections of \cite{zou_induced_1991,wang_induced_1991}}. The interferometric visibility is reduced if path-distinguishability is introduced via different transmissions \cite{zou_induced_1991,wang_induced_1991}, different temporal delays \cite{zou_control_1993} or different transverse sizes of the two idler beams \cite{barbosa_degree_1993,grayson_spatial_1994}.
Recently, the phenomenon led to applications in imaging \cite{lemos_quantum_2014}, spectroscopy \cite{kulik2004two,kalashnikov_infrared_2016}, metrology \cite{hudelist2014quantum}, spectrum shaping \cite{iskhakov2016nonlinear}, and in fundamental tests of complementarity (see e.g.  \cite{herzog_complementarity_1995,heuer_induced_2015,heuer_phase_2014}).

In this letter, we introduce and experimentally demonstrate a method for measuring the transverse momentum correlation between two photons by detecting only one of them. Our method is based on induced coherence without induced emission in a spatially multimode scenario.
We produce an interference pattern in the signal beams generated by two identical sources, which resembles fringes obtained in a Michelson interferometer. We show that the visibility of this pattern depends on the momentum correlation between signal and idler photons. We demonstrate how this can be used to obtain quantitative information about the momentum correlation.

\begin{figure}[b]
\includegraphics[width=\linewidth]{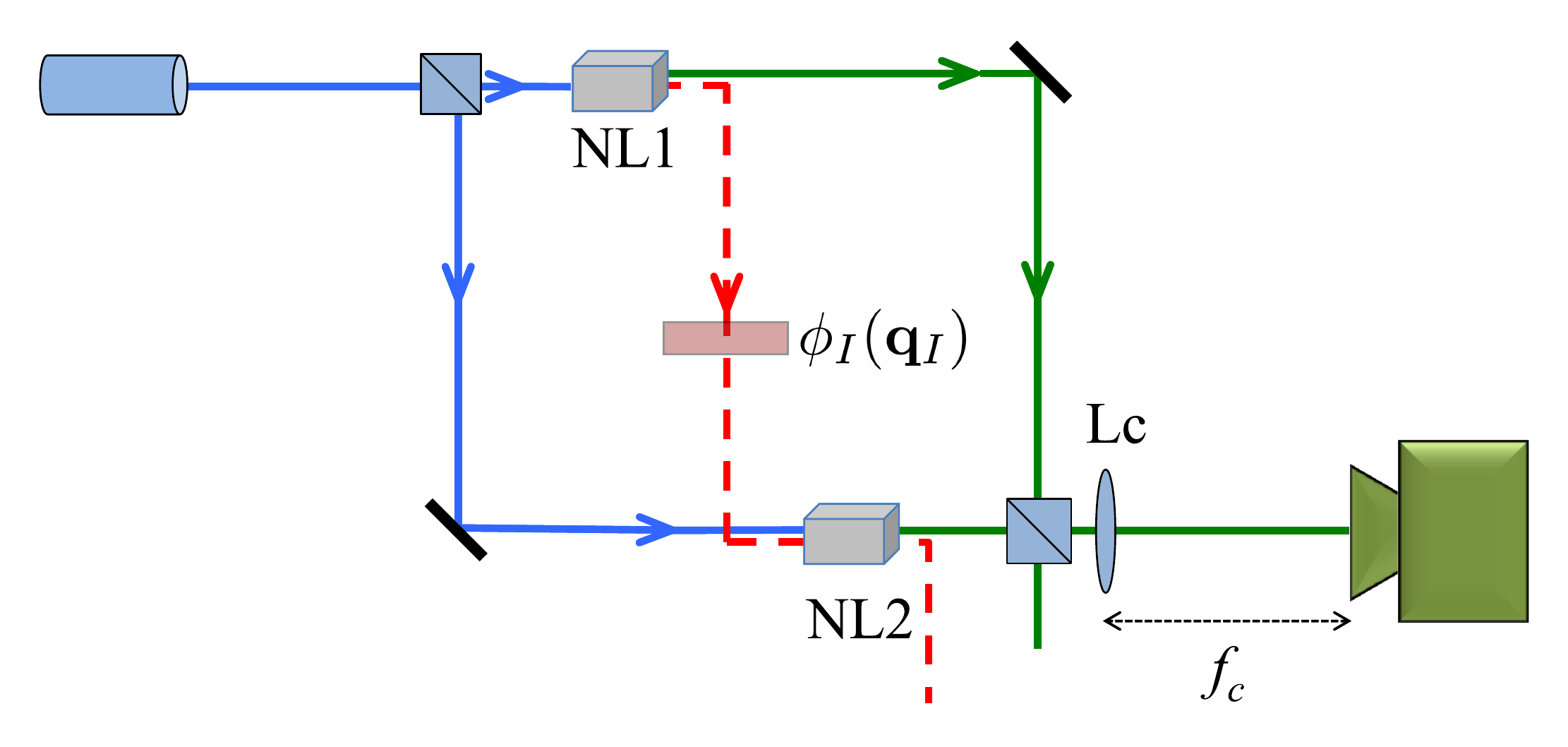}
\caption{Setup of the experiment. A continuous-wave pump laser (blue, 532 nm) is split at a beam-splitter and focused into two identical ppKTP crystals NL1 and NL2. The crystals produce signal (green, 810 nm) and idler (red, 1550 nm) photon pairs by type-0 SPDC. The idler beam from NL1 is overlapped with the idler beam from NL2 such that the two beams are indistinguishable after NL2. The two signal beams are superposed at a beam splitter and a camera detects the output at the focal distance of a positive lens (Lc).}
\label{fig:setup}
\end{figure}

In our experiment (Fig. \ref{fig:setup}), two identical periodically poled nonlinear crystals (NL1 and NL2) are pumped coherently ($\lambda_P$ = 532 nm) and emit photon pairs by collinear type-0 SPDC.
The idler beam ($\lambda_I$ = 1550 nm) generated by NL1 is directed through NL2 in order to overlap with the idler beam produced at NL2. After NL2, the idler beam is discarded. The two signal beams ($\lambda_S$  = 810 nm) are superposed at a beam-splitter. The path lengths are chosen such that the detection of a signal or an idler photon yields no information from which of the two crystals it had emerged.
As the pump power is low (150 mW), we can neglect the probability of creating more than one photon pair at a time and thus the effect of stimulated emission \cite{wiseman_induced_2000,lemos_quantum_2014}.

An EMCCD camera collects the superposed signal beam after it traverses an narrow-band (1 nm) 
frequency filter centered at 810 nm and a lens at focal distance from the camera. We assume the beams to be paraxial and the detection plane normal to the beam axis. In this case, the lens ensures that modes of the superposed signal beam with different transverse wave vectors $\qq_S$ are detected at distinct points $\mathbf{r_{\qq_S}}=\qq_S(f_c\lambda_S/2\pi)$ on the camera.

Confocal lens systems (not in figure) assure identical pump spots at NL1 and NL2, and point-by-point overlap of the idler beams. This causes each individual idler mode from NL1 to be indistinguishable from an equally populated idler mode at NL2 and coherence between the respective signal modes is induced \cite{lemos_quantum_2014,lahiri_theory_2015}.
An additional lens system images the signal beam at NL1 to the equivalent plane of NL2, canceling the spatially dependent phase shift that would arise due to the relative propagation distance between of two signal beams.
In this initial alignment position, the lens systems assure a spatially uniform interferometric phase between the two interfering signal beams. If, however, plane wave modes of the idler beam with different transverse components acquire different phase shifts $\phi_I(\qq_I )$ on their propagation from NL1 to NL2, the interference pattern in the superposed signal beam exhibits a spatial modulation.

It can be shown that the detected intensity pattern is given by (cf. \cite{fringetheory})
\begin{equation}
I(\mathbf{r_{\qq_S}})\propto p_S(\qq_S)\int p(\qq_I|\qq_S) \left(1+\cos[\phi_I(\qq_I)+\phi_0]\right) d\qq_I,
\label{cameraintensity}
\end{equation}
where $\phi_0$ contains all phase terms that are constant across the beam cross section.
Here, $p(\qq_I|\qq_S)=p(\qq_I,\qq_S)/p_S(\qq_S)$ describes the conditional probability density of detecting an idler photon with transverse momentum $\hbar\qq_I$ given its partner signal photon has transverse momentum $\hbar\qq_S$.
The variance $\Delta p(\qq_I|\qq_S)$ corresponds to the range over which the momentum of an idler photon can vary, when the momentum of its partner signal photon is known, i.e. to the ``width" of the correlation.

The visibility at each point on the interference pattern is evaluated by varying $\phi_0$ and computing $v(\mathbf{r_{\qq_S}})=[I_{max}(\mathbf{r_{\qq_S}})-I_{min}(\mathbf{r_{\qq_S}})]/[I_{max}(\mathbf{r_{\qq_S}})+I_{min}(\mathbf{r_{\qq_S}})]$. It follows from Eq. (\ref{cameraintensity}) that for a given $\phi_I (\qq_I)$, the visibility depends only on the conditional probability density $p(\qq_I|\qq_S)$.
Therefore, under reasonable assumptions, the measured visibility can be used to determine $\Delta p(\qq_I|\qq_S)$.

\begin{figure*}[htbp]
\includegraphics[width=\textwidth]{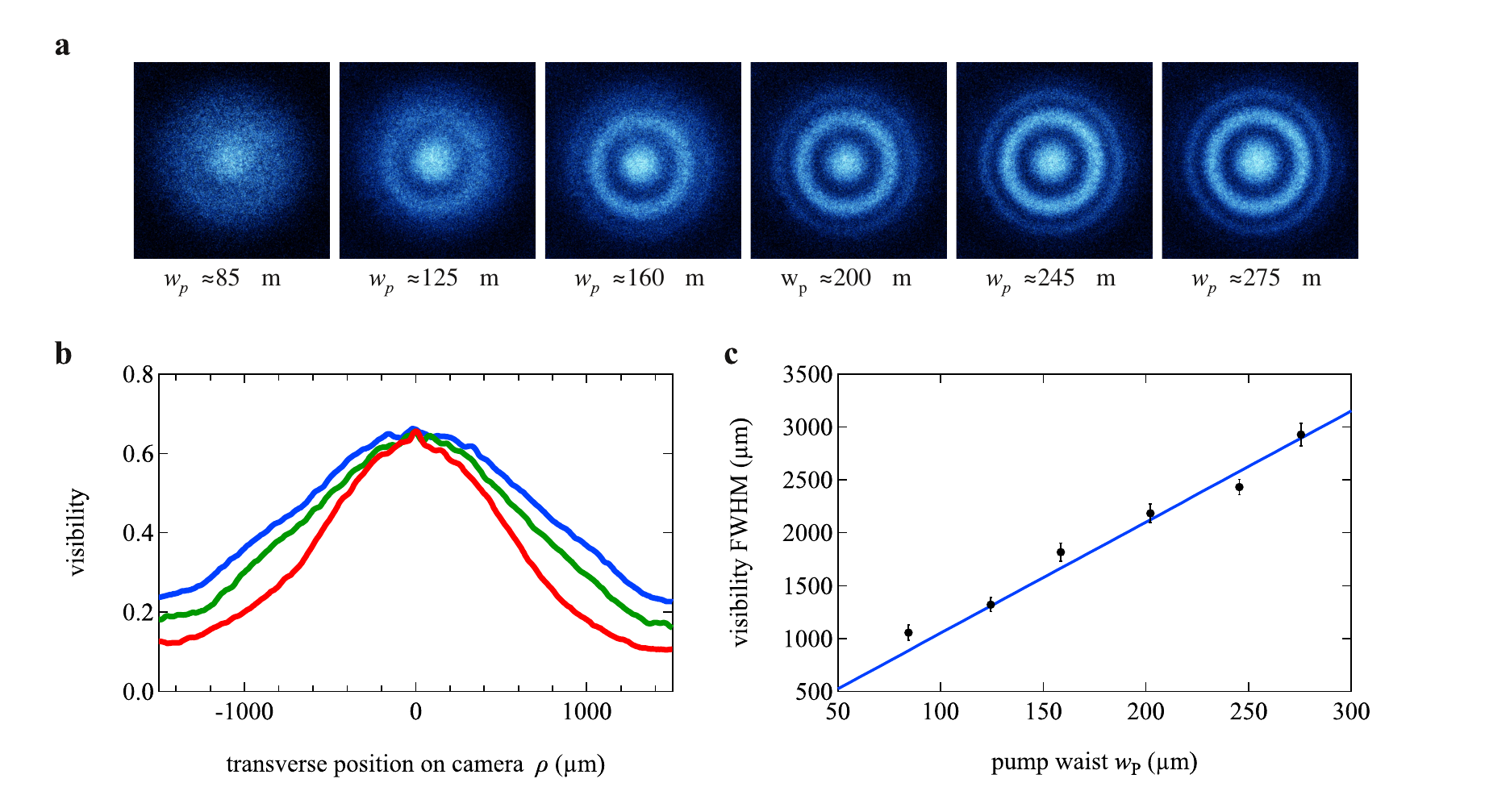}
\caption{Effect of a varying momentum correlation on the fringe visibility. Interference fringes are obtained in the signal beam by introducing an effective propagation distance ($d$ = 11.7 mm) in the idler beam between the two crystals. (a) Examples of resulting camera images for different Gaussian pump waists $w_P$. Smaller $w_P$ correspond to a weaker momentum correlation and to a lower visibility (b) Visibility profiles evaluated from the data by scanning the relative phase in which the two crystals are pumped, for $w_P\approx$ 125 $\mu$m (red), $w_P\approx$ 160 $\mu$m (green) and $w_P\approx$ 200 $\mu$m (blue). (c) Measured FWHM of the visibility as a function of the pump focus (black points) compared to a numerical simulation (blue line). 
}
\label{fig:visibility}
\end{figure*}

In order to analyze how the momentum correlation between signal and idler photons affects the visibility of the interference pattern, we first consider the case of perfectly anti-correlated transverse momenta, that is $p(\qq_I|\qq_S)\propto \delta(\qq_I+\qq_S)$. In this case, it follows from Eq. (\ref{cameraintensity}) that the intensity at a particular point on the camera depends on the phase introduced on exactly one $\qq_I$. As a result, the obtained interference pattern attains unit visibility.

On the other hand, if the momenta of signal and idler photons are uncorrelated, $p(\qq_I|\qq_S)$ 
is independent of $\qq_S$. As a result, the interference pattern described by Eq. (\ref{cameraintensity}) exhibits no spatial modulation. In this case, any variation of $\phi_I$ with $\qq_I$ leads to a vanishing visibility across the entire detected beam.

In general, the transverse momentum correlation between the two photons is neither perfect nor non-existent and $\Delta p(\qq_I|\qq_S)$ is finite.
It follows from Eq. (\ref{cameraintensity}) that in this case, the phases acquired by several $\qq_I$ contribute to the intensity modulation of the signal beam at one point on the camera. If  $\phi_I (\qq_I)$ varies within the range of contributing $\qq_I$, the camera detects an average over several interference patterns modulated by different phases.
As a consequence, the visibility of the resulting fringes is reduced compared to the perfectly correlated case. This reduction of visibility is generally more pronounced, the larger $\Delta p(\qq_I|\qq_S)$.

We introduced a spatially varying phase-shift by translating a lens of the imaging system in the undetected idler beam between NL1 and NL2 along the optical axis. This phase shift is to a good approximation equivalent to a free-space propagation about a distance $d$ \cite{fringesWL}. It is given by
\begin{equation}
\phi_I(\qq_I)=\frac{\lambda_I d}{4\pi}|\qq_I|^2,
\end{equation}
where the distance was set to $d$ = 11.7 mm in our experiment. The resulting interference pattern in the camera is circularly symmetric as the intensity value depends only on the radial distance from the beam center $\rho$. No spatially dependent phase was introduced in either of the detected signal beams.

The pump beam was focused equally into both crystals, subsequently using lenses of different focal lengths. The width of the momentum correlation was tuned by varying the size of the pump focus spots at NL1 and NL2 simultaneously (see e.g. \cite{monken_transfer_1998,walborn_spatial_2010}). A narrow focus corresponds to more pump wave vectors contributing to the creation of down-converted photon pairs. The resulting uncertainty in transverse momentum of the pump beam leads to a weaker momentum correlation between signal and idler photons compared to cases, in which the pump beam is closer to collimation. Figure \ref{fig:visibility}a shows experimentally obtained interference patterns for different Gaussian pump waists $w_p$. In each pump configuration, the interferometric phase $\phi_0$ was scanned.
The visibility at each pixel on the camera was evaluated from the image data (supplementary). The obtained visibility profiles express a maximum value at the center of the interference pattern and decrease with the distance from the center. This behavior is expected, as the introduced quadratic phase $\phi_I(\qq_I)$ varies faster for larger $\qq_I$. 
Figure \ref{fig:visibility}b shows that a narrower pump focus corresponds to a faster decrease. The full width at half maximum (FWHM) of each visibility profile was determined. The results are presented in Fig. \ref{fig:visibility}c in comparison to a numerical simulation.

\begin{figure}[b] 
\centering
\includegraphics[width=.45\textwidth]{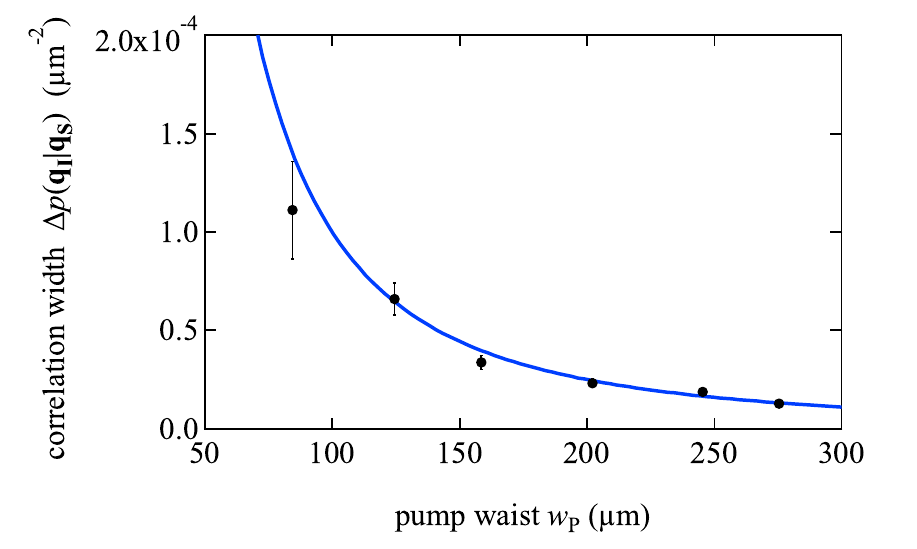}
\caption{Experimentally determined variance of the transverse momentum correlation between signal and idler photons ($\Delta p(\qq_I|\qq_S)=\sigma_c^2$) for different Gaussian pump waists in both crystals. The black data points are obtained from measurements on signal photons only. The results are compared to the theoretical prediction, $\Delta p(\qq_I|\qq_S)=1/w_p^2$ (blue line).
}
\label{fig:correlationWidth}
\end{figure}

In order to quantitatively relate our results to the variance of the transverse momentum correlation between signal and idler photons, the conditional probability density $p(\qq_I|\qq_S)$ needs to be parametrized. We assume that in the evaluated region of the beam, it can be approximated by a Gaussian distribution,
\begin{equation}
p(\qq_I|\qq_S) \propto \exp\left[-|\qq_S+\qq_I|^2/(2\sigma_c^2)\right],
\label{fittingmodel}
\end{equation}
where the standard deviation $\sigma_c$ represents a measure for how well the transverse momenta of signal and idler photons are correlated. This parameter is determined experimentally. Equation (\ref{fittingmodel}) can be justified by comparison to the theoretical prediction for the SPDC emission in our experiment (supplementary). The resulting visibility can be explicitly calculated using Eq. (\ref{cameraintensity}).

A larger value of $\sigma_c$ corresponds to a faster decrease of the visibility with the distance from the center. It can be shown that a one-to-one correspondence exists between the FWHM of the peak of the visibility profile and $\sigma_c$ \footnote{An explicit formula for the visibility in this situation is given by Eq. (22) of \cite{fringetheory}. There, the quantity $\sigma_\theta$ is proportional to $\sigma_c$ used in our treatment. It follows from this formula that the one-to-one correspondence holds in the regime of our interest.}. This fact allows to numerically compute the variance of the conditional transverse momentum distribution 
$\Delta p(\qq_I|\qq_S) = \sigma_c^2$ from the measured FWHM values. Figure \ref{fig:correlationWidth} shows the results in comparison to the theoretical predictions.

In our approach, we circumvented the necessity of evaluating absolute visibility values, which is sensitive to experimental imperfections and noise \footnote{In fact the absolute visibility at the center of the beam changes only minutely for the different values of $w_P$ in our experiment. Moreover, the absolute visibility is very sensitive to alignment stability. We experimentally quantified the correlation via the FWHM, as this method relies only on relative visibilities and allows to robustly distinguish the obtained visibility profiles.
}. Instead, we determined $\Delta p(\qq_I|\qq_S)$ merely from relative visibility values obtained at different radial distances from the beam center on the camera.

We have demonstrated a method of quantifying the transverse momentum correlation between signal and idler photons by detecting only signal photons.
Our method requires two sources, which have identical emission properties and emit photon pairs coherently. The latter is a necessary condition for the effect of induced coherence without induced emission to be observed. We assumed a specific Gaussian form of the conditional probability density of signal and idler transverse momenta. However, our approach could be easily extended to different distributions.

Experimentally, we determined the momentum correlation between two photons independently of their correlation in position. All measurements were performed only on signal photons and only in the momentum basis. We did not need to measure the spatial coherence of the signal beam 
\footnote{Measurements of the spatial coherence in one beam have been used previously to show entanglement of bi-photons under the assumption of pure states (see e.g. \cite{abouraddy_demonstration_2001,pires_direct_2009,just_transverse_2013}).
}. Furthermore, we did not perform any coincidence detection or post selection.

The presented method might be used for the characterization of sources of momentum correlated photons in situations where conventional methods involving coincidence detection are not practical, e.g. if detectors are not available for both wavelengths. From a fundamental viewpoint, it demonstrates that it is possible to measure higher order correlations between two systems without detecting one of them.

We believe that our method can be generalized to measure correlations not only in transverse momentum but also in other degrees of freedom, particularly in transverse position. This potentially has far-reaching implications in the characterization of spatially entangled states, as it opens a promising avenue of research towards an experimental measure of continuous-variable entanglement, which does not require a pure state assumption and which requires the detection of only one of two entangled particles.

\section*{Acknowledgements}
The authors thank F. Steinlechner for helpful discussions. This work was supported by the Austrian Academy of Sciences (\"OAW) - IQOQI Vienna and the Austrian Science Fund (FWF) with SFB F40 (FOQUS) and W1210-2 (CoQuS).

\let\Section\section 
\def\section*#1{\Section{#1}} 
\bibliographystyle{unsrt}

\clearpage
\subsection{Supplementary Information}
\subsubsection{Theory of SPDC correlations}
The bi-photon state produced in each nonlinear crystal can be written in the angular spectrum representation as (e.g. \cite{monken_transfer_1998})
\begin{equation}
\ket{\Psi}=\int C(\qq_S,\qq_I)\ket{\qq_S}\ket{\qq_I} d\qq_S d\qq_I,
\label{state_C}
\end{equation}
where $\qq_S$ and $\qq_I$ represent transverse components of a wave vector of the signal and the idler beams respectively. 
The the joint probability density of detecting an idler photon in mode $\qq_I$ and a signal photon in mode $\qq_S$ is given by $p(\qq_I,\qq_S) = |C(\qq_S,\qq_I)|^2$.
The spatial properties of the bi-photon emission are governed by (cf. e.g. \cite{monken_transfer_1998,grice_spatial_2011}) 
\begin{equation}
C(\qq_S,\qq_I)\propto A\left(|\qq_S+\qq_I|^2\right)\sinc{\frac{L \Delta k_z(\qq_S,\qq_I)}{2}},
\label{monkenmodel}
\end{equation}
where $A\left(|\qq_S+\qq_I|^2\right)=A\left(|\qq_P|^2\right)$ represents the angular spectrum of the pump beam, $L$ is the length of the nonlinear crystal and  $\Delta k_z(\qq_S,\qq_I)$ is determined by the phase matching.
In the absence of transverse phase mismatch, 
\begin{equation}
\Delta k_z(\qq_S,\qq_I)=|\qq_S-(\lambda_I/\lambda_S)\qq_I|^2\frac{\lambda_P\lambda_S}{4\pi\lambda_I},
\label{longphmismatch}
\end{equation}  
where $\lambda_S$, $\lambda_I$ and $\lambda_P$ denote the wavelengths of signal, idler and pump beams. Equation (\ref{longphmismatch}) representing the longitudinal phase-mismatch $\Delta_z$ was obtained by generalizing the derivation in \cite{grice_spatial_2011} to the case of non-degenerate SPDC. We implicitly assume the material contribution to the phase-mismatch to be much smaller than the geometrical contribution, which holds for most optical materials \cite{grice_spatial_2011}.

If the pump is a Gaussian beam with waist $w_P$, its angular spectrum is given by
\begin{equation}
A\left(|\qq_S+\qq_I|^2\right)=\exp\left(-|\qq_S+\qq_I|^2 w_P^2/4\right).
\label{pumpangspec}
\end{equation} 
Equation (\ref{pumpangspec}) accounts for the fact that a narrower pump focus leads to weaker momentum correlation due to a higher uncertainty in transverse momentum of the pump photons \cite{monken_transfer_1998,tasca_propagation_2009,walborn_spatial_2010}. In our experiment, we evaluate the visibility in a circular region around the beam center of the superposed signal beam of $\approx$ 2 mm radius, corresponding to $|\qq_S|\approx10^5$m$^{-1}$. Eq. (\ref{longphmismatch}) was computed four our crystal parameters and was found to be almost constant within this range. Therefore, the bi-photon amplitude in Eq. (\ref{monkenmodel}) can be approximated by a Gaussian function.

It follows that the conditional probability density $p(\qq_I|\qq_S)=|C(\qq_S,\qq_I)|^2/\int |C(\qq_S,\qq_I)|^2 d\qq_S$ is given by
\begin{equation}
p(\qq_I|\qq_S)\propto\exp\left(-|\qq_S+\qq_I|^2 w_P^2/2\right),
\end{equation}
which corresponds to Eq. (\ref{fittingmodel}), with a predicted variance $\sigma_{theo}^2=1/w_p^2$. This expression was used to perform the numerical simulations.\\

\subsubsection{Evaluation of the visibility FWHM values}

For each measurement, the raw camera images for 25 different settings of interferometric phase $\phi_0$ were corrected for background and a Gaussian image filter was applied to reduce noise.
Camera pixels, at which the mean measured intensity (averaged over all phase values) was lower than 20\% of the average intensity in the beam center were discarded. Independently for each camera pixel, sinusoidal fits were applied to the measured intensity values at different phase settings. The visibility at each point was computed from the resulting fit coefficients.
The visibility profile was obtained as an average over 201 cross sections evaluated at different angles. The center of symmetry was found by maximizing the cross correlation of the radial visibility profile in opposite directions.

\end{document}